\documentclass[10pt]{article}

\usepackage[utf8]{inputenc}
\usepackage[dvipsnames]{xcolor}
\usepackage{bbm}
\pdfoutput=24
\usepackage{amsmath}
\usepackage{float}

\usepackage[linesnumbered,ruled]{algorithm2e}
\usepackage{array} 
\usepackage{esvect}
\usepackage[toc,page]{appendix}
\usepackage{leftidx}
\usepackage{color}	
\usepackage{framed, color}
\usepackage{multirow}
\usepackage{pdfpages}
\usepackage{multicol}
\usepackage{wrapfig,lipsum,booktabs}
\usepackage{relsize}
\usepackage{mathtools,hyperref}
\usepackage{xparse}
\usepackage{amsthm}

\hypersetup{
    colorlinks=true,
    linkcolor=cyan,
    filecolor=cyan,
    urlcolor=red,
    citecolor=red
}

\usepackage{cleveref}
\usepackage{commath}
\usepackage{amssymb}

\crefformat{section}{\S#2#1#3} 
\crefformat{subsection}{\S#2#1#3}
\crefformat{subsubsection}{\S#2#1#3}

\definecolor{mgreen}{RGB}{25,147,100}
\definecolor{shadecolor}{rgb}{1,.8,.1}
\definecolor{shadecolor2}{RGB}{245,237,0}
\definecolor{orange}{RGB}{255,137,20}
\definecolor{orange}{RGB}{245,37,100}

\usepackage{pgfplots}
\usetikzlibrary{patterns}
\usepackage{mdframed}
\usepackage{adjustbox}
\usepackage{tcolorbox}
\usepackage{graphics}
\usepackage{tikz,ifthen,fp,calc}
\usepackage{caption}
\usepackage{subcaption}
\usetikzlibrary{plotmarks}

\theoremstyle{plain}

\theoremstyle{definition}

\theoremstyle{definition}

\theoremstyle{example}

\usepackage[english]{babel}
\usepackage{babel,blindtext}

\usepackage{fullpage}
\usepackage{amsfonts}
\usepackage{lscape}
\usepackage{bbm}

\usepackage{todonotes}
\usepackage{verbatim}
\usepackage{bm}

\usepackage[margin=.5in]{geometry}
\providecommand{\keywords}[1]{\textbf{\textit{Keywords:---}} #1}

\usepackage[T1]{fontenc}
\usepackage[utf8]{inputenc}
\usepackage{authblk}
\usepackage{cite}

\title{\textbf{Recent advances in opinion propagation dynamics: A 2020 Survey}}

\author[1]{Hossein Noorazar\thanks{hnoorazar@math.wsu.edu}}
\affil[1]{Washington State University, Pullman, Washington, United States of America}
\date{}

\usepackage{graphicx}

\providecommand{\keywords}[1]{\textbf{\textit{Keywords:}} #1}

\usepackage{siunitx}
\begin{document}
\maketitle

\abstract{
Opinion dynamics have attracted the interest of researchers from different fields.
Local interactions among individuals create interesting dynamics
for the system as a whole.
Such dynamics are important from a variety of perspectives. 
Group decision making, successful marketing, and 
constructing  networks  (in which consensus can be reached or prevented) 
are a few examples of existing or potential applications.
The invention of the Internet has made the opinion fusion faster, unilateral, and on
a whole different scale. Spread of fake news, 
propaganda, and election interferences have made it clear there is an essential 
need to know more about these dynamics.

The emergence of new ideas in the field has accelerated 
over the last few years. In the first quarter of 2020, at least 50
research papers have emerged, either peer-reviewed and published
or on pre-print outlets such as arXiv. In this paper,
we summarize these ground-breaking ideas and their fascinating extensions
and introduce newly surfaced concepts.
}


\tableofcontents
\begin{multicols}{2}
\section{Introduction}\label{intro}
Opinion dynamics studies propagation of opinions in a network through 
interactions of its agents. 
Modeling opinion dynamics goes back a few decades.
Asch in 1951~\cite{Asch1951} studied the effect of group pressure
on social dynamics.
French in 1956 was among the first researchers to devote attention to opinion dynamics;
\emph{A Formal Theory of Social Power}~\cite{FRENCH1956}. In 1964
Abelson~\cite{abelson1964mathematical} proposed a continuous-time model.
A decade later, in 1974, DeGroot established one of the 
simplest models~\cite{DeGroot1974}, which has become one of the most well known. 
Two years later Lehrer~\cite{lehrer1976rational} also developed his model
that is identical to that of DeGroot and an extensive discussion by Lehrer and Wagner 
are given in~\cite{lehrer2012rational}.
Another pioneer and interesting work is~\cite{latane1981psychology} by Latan{\'e} where
he explored the interactions from a psychology angle and reviewed some
of the earlier works such as group pressure, immitation, and effects of newspapers.

Opinion dynamics take different shapes depending on the nature of the 
topic under consideration and the purpose of interactions.
For example, the topic
could be liking or disliking a certain food such as fish; here,
binary opinion dynamics comes into play~\cite{Biswas2009,Ding20101,mukhopadhyay2020}. 
Sometimes the opinion can be represented by continuous variables. 
For example, the extent to which one supports a cause.
Over the years, a few different models for 
continuous-opinions have been 
proposed~\cite{FRENCH1956,Deffuant2000,Noorazar2016}.
Either as an abstract idea or in real life, one can think of a continuous-opinion space
in which one must take discrete actions~\cite{Martins2007,Martins2020Discrete}. 
For instance, in an election
where each agent's support for candidates falls on the continuous spectrum,
each agent must still cast a discrete vote.

In short, opinion dynamics can be  explained as follows.
Agents start with an initial opinion. Connected agents
interact and update their opinion by a given clear update rule.
This process is carried on until a termination criterion is met.
An example of termination criteria is reaching a steady state
in which agents do not change their opinion anymore.

In this paper, we briefly introduce the main concepts and
newly developed  ideas of modeling opinion dynamics. 
Going into details is not the purpose of this survey.
In Sec.~\ref{SecMilestones} the major models are presented,
then, in Sec.~\ref{SecMilestonesExten} basic results and certain extensions
of the well-known models are reviewed. In Sec.~\ref{LastWords}, we go through
some models that have not been studied 
exhaustively but are interesting and
have contributed novel concepts. 
Finally, the conclusions are presented
in Sec.~\ref{SecConcl}.

\section{Preliminaries}
The network of agents is denoted by $\mathcal{G}$ in which $N$ agents are present. 
Let $\mathbf{A}$ be the adjacency matrix
where $\mathbf{A}_{ij} = 1$ if agents $i$ and $j$ are connected and $\mathbf{A}_{ij} = 0$ otherwise.
The row-stochastic \emph{influence matrix} is denoted by $\mathbf{W}$
where $\mathbf{W}_{ij}$ is the level of influence of agent $j$ on agent $i$; $0 \le \mathbf{W}_{ij} \le 1$.

Let us define the opinion space to be the set of
all possible opinions denoted by $\mathcal{O}$. 
Examples of opinion space are $\mathcal{O}= \{0, 1\}, \{1,2 , \dots m\}, [0, 1]$. 
The opinion of agent $i$ at time $t$ is denoted by $o_i^{(t)}$.
The state of the system at time $t$ is denoted by
$ \mathbf{o}^{(t)} = \begin{bsmallmatrix} o_1^{(t)} & o_2^{(t)} & \cdots & o_N^{(t)} \end{bsmallmatrix}$.

When the system reaches equilibrium, we say the
system has converged. The convergence state may be
consensus, polarization, or fragmentation. Polarization
is the state in which there are only two clusters of agents, and  
fragmentation is the state there are more than two clusters.

Before moving to the next section, we would like 
to mention that there is no convergence on terminology in the literature.
For example, consider an agent that does not change its opinion
over time; some papers refer to such an agent 
as a \emph{leader}~\cite{yi2019disagreement}, 
others as a \emph{media}~\cite{brooks2019model}, 
some as an \emph{stubborn agent}~\cite{abrahamsson2019} 
and a few as an \emph{inflexible agent}~\cite{GalamInflex,Galaminflexible}. 
A \emph{closed-minded} agent is referred 
to someone who does not change its opinion in~\cite{Chazelle2017} 
and in some papers, a closed-minded agent is an agent whose 
confidence radius is small compared to other agents~\cite{lorenz2010heterogeneous}. 
We adhere to the following definitions. An agent that does 
not change its opinion over time is referred to as a 
\emph{fully-stubborn} agent. An agent that weighs its initial 
opinion, i.e., takes its initial opinion into account, in all 
interactions over time is referred to as a \emph{partially-stubborn} 
agent. If an agent is not stubborn, it is called a \emph{non-stubborn} 
agent. In the bounded confidence models, we say an agent 
is closed-minded if its confidence radius is smaller than that of others.
\section{Milestones}\label{SecMilestones}
In this section, we present the main models that have inspired a
tremendous amount of research. We start with the models in which
the opinion space is continuous and then present the models
in which the opinion space is discrete.
\subsection{Continuous opinion space models}\label{contOpModels}
In this section, we present the DeGroot model and its major extension
known as the Friedkin-Johnsen model. 
Next, we examine the bounded confidence models in which
agents interact with those whose opinions are close enough to that
of their own.
\subsubsection{DeGrootian models}\label{DeGrootian}
We begin with the simplest model, the DeGroot model.
Moreover, since the Friedkin-Johnsen extension of the DeGroot model is well-known
and has been studied extensively, we present it here as well.

\begin{figure*}
\centering
\includegraphics[width=.35\linewidth]{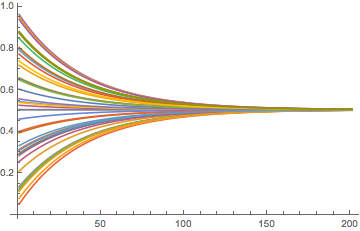}
\caption{Evolution of the DeGroot model from initial profile to consensus.}
\label{fig:degrootevolution}
\end{figure*}

\textbf{DeGroot model.} The DeGroot model~\cite{DeGroot1974} is given by
\begin{equation}\label{eq:DeGroot}
\mathbf{o}^{(t)} = \mathbf{W o}^{(t-1)} = \mathbf{W}^2 \mathbf{o}^{(t-2)}  = \cdots = \mathbf{W}^t \mathbf{o}^{(0)}
\end{equation}
where $\mathbf{W}$ is a \emph{row-stochastic} weight matrix and 
$\mathbf{W}^t$ is its $t$-th power.
The model is linear and traceable over time. 
Classical linear algebra tools are sufficient to analyze this model. It has been very well studied
and different extensions of it exist. The DeGroot model is an iterative averaging model. 
If the network is connected then convergence is equivalent to consensus (Fig.~\ref{fig:degrootevolution}).
Berger~\cite{Berger1980} showed the DeGroot model will 
reach consensus if and only 
if there exists a power $t$ of the weight matrix for which 
$ \mathbf{W}^t$ has a strictly positive column.
Let $\mathbf{W}$ be such a matrix; then, the consensus opinion is 
given by $\label{eq:lefteig} o^* = \langle \bm{\ell}_\mathbf{A},  \mathbf{o}^{(0)}  \rangle $
where $ \bm{\ell_W^\text{T}}$ is the left eigenvector of $\mathbf{W}$ 
associated with 1,  
constrained to $\langle \bm{1}_N, \bm{\ell_A} \rangle = 1$.

\textbf{Friedkin-Johnsen model.}
One of the major extensions of the DeGroot model was introduced by
Friedkin and Johnsen~\cite{FJmodel1990,FJmodel} and is known as the Friedkin-Johnsen (FJ) model. 
Since it has functioned as a ground-breaking
model, we include it here instead of in Sec.~\ref{SecMilestonesExten}. 
In the FJ model, the idea of \emph{stubborn-agents} is added to the DeGroot model:
\begin{equation}\label{eq:FJmodelEq}
\mathbf{o}^{(t+1)} = \mathbf{D} \mathbf{W o}^{(t)} + ( \mathbf{I} -  \mathbf{D}) \mathbf{o}^{(0)}
\end{equation}
\noindent where $\mathbf{D} = diag([d_1, d_2, \dots, d_N])$ 
with entries that specify the \emph{susceptibility} of individual 
agents to influence, i.e., $(1-d_i)$ is the level of 
stubbornness of agent $i$. For a fully-stubborn agent 
$1-d_i$=1, for a partially-stubborn agent $0 < 1-d_i < 1$
and for a non-stubborn agent $1-d_i = 0$.
$\mathbf{W}$ is a row stochastic influence matrix. The convergence and 
stability of the FJ model are studied in~\cite{Parsegov7577815}.\\

These two models are the main two DeGrootian models.
We now move on to the bounded confidence models.
\subsubsection{Bounded confidence models}\label{BCMsec}
In this section, we look at bounded confidence models.
A bounded confidence model (BCM) is a model in which 
agents ignore the ideas that are too far from 
their own. The well-known pairwise BC model 
is given by Deffuant et al.~\cite{Deffuant2000} and is called the DW model, while 
the most well-known synchronous version is given by
Hegselmann and Krause~\cite{Hegselmann2002}, and is called the HK model.

\textbf{Deffuant-Weisbuch model.}
The celebrated DW model of Deffuant et al.~\cite{Deffuant2000}
is defined by the following rule:
\begin{equation}\label{eq:BCM}
  \begin{cases} 
  o_i^{(t+1)}  = o_i^{(t)} + \mu \: . \: (o_j^{(t)} - o_i^{(t)})  \\
  o_j^{(t+1)}  = o_j^{(t)} + \mu \: . \: (o_i^{(t)} - o_j^{(t)})  \\
\end{cases}
\end{equation}
\noindent where $\mu$ is the so-called learning rate that usually 
lies in $(0, 0.5]$ to avoid crossover. The update takes place
only if $|o_j^{(t)} - o_i^{(t)}| \leq r$ where $r$ is called \emph{confidence radius}.
In~\cite{Deffuant2000} all agents share the same confidence radius $r$
and the same learning rate $\mu$. Said differently, the system
is homogeneous in both $r$ and $\mu$. Obvious variations
can be achieved by introducing a heterogeneous confidence 
radius to the system, adding asymmetry in the 
confidence radius, or even an agent-specific 
time-varying confidence radius~\cite{Sobkowicz}.
Intuitively and in actuality, the confidence radius affects the
number of clusters at equilibrium. Consider the 
opinion space $\mathcal{O} = [0, 1]$ and confidence radius $r$
to be 1. Then, each agent can interact with any other agent at all times, and the
subspace in which agents lie in will be contractive. 
In another scenario, let $r < m = \min_{i,j}\{|o_i^{(0)} - o_j^{(0)}|\}$;
then, there will be no interaction, and thus, there will be 
$N$ clusters of size 1.
In fact, Fortunato~\cite{Fortunato2004} claims $r=0.5$ is the
critical confidence radius above which the agents come to a
consensus. Moreover, the number of clusters at equilibrium is
approximately $1/2r$~\cite{castellano2009statistical}. 
Lorenz~\cite{lorenz2010heterogeneous} investigates
a heterogeneous (in confidence radius) case in which there are two groups
of agents. One group is closed-minded, i.e., has a smaller confidence radius,
and the other group is open-minded.
It is shown in~\cite{lorenz2010heterogeneous} 
that heterogeneity of confidence radius helps consensus to be reached;
the final state will be a consensus when the confidence radius 
of the open-minded group is well below the aforementioned 
$r = 0.5$ for the homogeneous case. Chen et al.~\cite{Chen2019} 
investigate convergence properties of a heterogeneous 
(in confidence interval) DW model. An asymmetric DW 
model is discussed in~\cite{zhang2015conv}. 
Shang~\cite{Shang2014} has proposed a modified DW 
model where confidence radiuses are assigned to 
edges, as opposed to agents. Equivalently, 
agent $i$ trusts agent $j$ and $k$ differently; 
the convergence properties of such a model
are studied by Shang~\cite{Shang2014}. Another work that studies the convergence
properties of a modified DW is ~\cite{Huang2018},
in which the learning rate is a function of opinion difference.

\textbf{Hegselmann-Krause model.}
The most well-known synchronous version of BCM is given
by Hegselmann-Krause~\cite{Hegselmann2002}. 
The (most common and simplest) update rule of the HK model is given by:
\begin{equation}\label{eq:HKmodel}
o_i^{(t+1)} = \frac{1}{|N_i^{(t)}|} \sum_{j \in N_i^{(t)} } o_j^{(t)}
\end{equation}
\noindent where $N_i^{(t)}$ is the set of neighbors of
agent $i$ at time $t$, i.e., the set of agents for whom we have
$| o_j^{(t)} - o_i^{(t)} | \leq r $, including $i$ itself. 
Reference~\cite{Hegselmann2002} includes
the analyses of convergence and consensus for the HK model.
Bhattacharyya et al.~\cite{Bhattacharyya} study convergence
properties of a multidimensional HK model.
\subsection{Discrete opinion space models}\label{DiscOpModels}
In this section, we focus on opinion models whose opinion space
is discrete. Variations of the Ising model provide examples with 
binary opinion space. A discretized version
of DW~\cite{Dietrich2003} is another example. These models have applications
in real life. 
At least two studies used binary opinion models to explain 
Trump's 2016 victory~\cite{Galam2017,Biswas2017}.
\subsubsection{Galam model}
In addition to Friedkin, who has left a large footprint in this field since 1986~\cite{friedkin1986formal},
Galam has spent more than 35 years studying opinion dynamics 
from a sociophysics perspective~\cite{galam1982,GALAM1986426}  
and his work has inspired other researchers. 
He has studied a range of different dynamics~\cite{GalamReview} including 
``democratic voting in bottom-up hierarchical systems, 
decision making, fragmentation versus coalitions, terrorism and opinion dynamics.'' 
Reference~\cite{GalamReview} reviews Galam's work prior to 
2008 and further details can be found in his 
book~\cite{GalamBook}. 
Here, we introduce some of the 
newer works related to the binary opinion space used in the Galam model.

In the Galam model, there are two opinions in the opinion space. The update rule
is as follows;
(1) agents are randomly distributed in groups of size $r$, 
(2) each group uses majority rule to update their opinion, then
(3) agents are shuffled and the cycle begins again at step (1).

G{\"a}rtner and Zehmakan~\cite{GortnerBernd}
address consensus time and sensitivity of outcome 
as functions of initial state in the Galam model.
\subsubsection{Sznajd model}
Ising models have a long history in statistical physics. Here, we
overview one of the well-known models of this kind in the
field of opinion dynamics, namely the Sznajd model~\cite{Sznajd2000}. 
In the  Sznajd model, $N$ agents are sitting on a 1-dimensional lattice.
Opinion space is given by $\mathcal{O} = \{-1, + 1\}$.
At a given time $t$, two neighbors $i$ and $i+1$ are selected randomly.
If $o_i^{(t)} \times o_{i+1}^{(t)} = 1$ then agents $i-1$ and $i+2$ adopt the
direction of agents $i$ and $i+1$, otherwise, the agent $i-1$ 
adopts the opinion of agent $i$ and agent $i+2$ adopts the opinion of
its selected neighbor, agent $i+1$. Steady states of such a model
have all agents in agreement at either +1 or -1 or a stalemate. The
time needed to reach equilibrium is discussed in~\cite{Sznajd2000} through 
Monte Carlo simulations. Some results from the original Sznajd model
and the Sznajd model on a complete graph are presented in~\cite{slanina2003analytical}. 

Phase transition phenomena in the Sznajd model with the presence of anticonformists
in complete graphs are examined by~\cite{MuslimRoni,Calvelli}. 
Calvelli et al.~\cite{Calvelli} also consider 2D and 3D lattices.
To learn more about the Sznajd model please see~\cite{sznajd2005sznajd,STAUFFER200293}
\subsubsection{Voter model}
In a voter model, opinion space is binary. At a given time $t$,
a random agent, $i$, is chosen. Then, $i$ chooses a random neighbor
and adopts the state of the neighbor.

The voter model on regular lattices has been studied extensively. 
There are also variations of the voter model on different network topologies.
 Sood and Redner and ~\cite{sood2005voter} investigate the voter model on a 
heterogeneous graph, and~\cite{Gastner2019} explores the dynamics and 
convergence time of the 
voter model on a graph with two cliques. 
The influence of an external source is investigated in~\cite{Majmudar}. 
The impact
of ``active links'' on the convergence of the voter model is investigated in~\cite{caridi2019topological}.
To learn more about voter models please 
see~\cite{REDNER2019275}. Examples of the other 
extensions are given below.
\section{Milestones' extensions}\label{SecMilestonesExten}
We are ready to investigate some of the fascinating extensions of the reviewed models in the previous section.
\subsection{Stubborn agents}
\textbf{Stubborn agents in DeGroot model.}
One major modification of the DeGroot model known as the FJ model 
 adds stubbornness and was presented
previously. However, here we introduce other versions that are new and 
have not studied extensively. Abrahamsson et al.~\cite{abrahamsson2019} 
study the effect of the presence of fully-stubborn agents in the DeGroot model.
Wai et al.~\cite{Wai2016} propose 
``an active sensing method to estimate the relative weight 
(or trust) agents place on their neighbors’''
and explore the role of stubborn agents in such an environment.
Zhou et al.~\cite{ZHOU2020363} study the effect of partially-stubborn
agents on a modified DeGroot model. In their altered model,
an agent not only takes the opinions of its neighbors into account but also
takes the opinions of its neighbors' neighbors into account as well.

\textbf{Stubborn agents in DW and HK.} 
The effect of 
Stubborn agents in DW and HK is explored in~\cite{huang2016modeling}
and~\cite{brooks2019model}, respectively. In the latter, stubborn agents
are labeled as media. They investigate ``how the number of media accounts and the number
of followers per media account affect the media impact.''
Moreover, one of their novel contributions is ``content quality.'' 

\textbf{Stubborn agents in Galam model.}
References~\cite{GalamInflex,CHEON20181509}
study the effect of \emph{inflexible} agents in the Galam model.
Another work~\cite{galam2019tipping} is an extension of the Galam model
in which they study how the minority wins against the majority in scenarios
such as the US election in 2016. The model given in~\cite{QIAN2015187} also has the
stubbornness ingredient where stubborn agents are called leaders and 
their power of influence is discussed. Cheon and Morimoto~\cite{CHEON2016429} consider a Galam model
that includes \emph{balancer} agents who oppose stubborn agents.
Contrarian agents are specific to the Galam model, hence, we include them 
here. Galam and Cheon~\cite{GalamAsymmetric} investigate the
effect of asymmetry in contrarian behavior which is an extension of~\cite{GalamInflex}.

\textbf{Stubborn agents in voter model.}
References~\cite{Mobilia2007} and~\cite{PhysRevEKhalil} explore the role 
of stubborn agents in a voter model and a noisy voter model, respectively. 
Yildiz et al.~\cite{yildiz2013binary}
examine the effect of stubborn agents with opposing views  on the convergence of
the system. Mukhopadhyay et al.~\cite{mukhopadhyay2020}
investigate the effect of biased agents in both voter and majority-rule
dynamics. They also add stubbornness to the majority-rule case.
The bias and stubbornness are implemented by
updating probability. 
They study the relationship between the size of the network
and (1) consensus time, and (2) probability of consensus. 
\subsection{Biased agents}
Biased agents are more open to agents' that hold similar 
opinions to themselves as opposed to others, i.e., the homophily quality. 

\textbf{Biased agents in DeGroot model.}
Dandekar et al.~\cite{Dandekar5791} incorporate
the idea of biased agents in the DeGroot model and turn it into
a nonlinear model. Xia et al.~\cite{xia2019analysis} provide some analysis 
for equilibria in such a model.

\textbf{Biased agents in DW model.}
In the DW model, a pair of agents are chosen 
randomly.  S{\^{i}}rbu et al.~\cite{Sirbu2019} 
modify the DW model to add the bias ingredient. In this model, agent $i$ is chosen 
randomly and then the interaction partner $j$ is 
chosen by a probability function that depends on the
difference of opinion. The closer the opinion of 
agent $j$ is to the opinion of $i$, the more the 
probability of interaction between them. 
Convergence properties and network size 
effects are addressed by~\cite{Sirbu2019}.

\textbf{Biased agents in HK model.}
Chen et al.~\cite{Chen2017} take the modified HK 
model of Fu et al.~\cite{Fu2015}
and extend it to include biased agents. They call the new 
model the ``Social-Similarity-Based HK model.''
In this model, for two agents to interact not only do they need
to hold close opinions but the criteria of social similarity also must be met, i.e.,
their other attributes need to be close as well.
Social similarity can be measured by considering different attributes
such as age, education, and other traits.
\subsection{Opinion manipulation}
Opinion manipulation is fascinating for different reasons.
Maximizing the number of customers in a market or interfering 
with another country's election are two examples
of opinion manipulation. Below, we review some of the proposed models.

\textbf{Opinion manipulation in DeGroot model.}
There are a few works about how to make
a network reach a consensus and further still, how to influence
 the agents toward a predetermined 
opinion~\cite{Dong2017,Pineda2015,Bauso2018}.
Dong et al.~\cite{Dong2017} propose a network modification, 
i.e., adding a minimum number of edges to the network, to reach consensus in 
the DeGroot model. Zhou et al.~\cite{ZHOU2020363} consider the manipulation of
public opinion in a modified DeGroot model.
Hegselmann et al.~\cite{hegselmann2014optimal} use one strategic 
agent who can change its opinion freely to steer as many as possible
agents towards a predetermined interval.

\textbf{Opinion manipulation in FJ model.}
To the best of our knowledge, there is no study to influence agents
toward consensus in the FJ model. 
Previously, a missing piece for manipulation of agents in models
was preventing a network from reaching a consensus. 
A very recent study, Gaitonde et al.~\cite{gaitonde2020}, investigates 
adversarial manipulation of a network to prevent it
from consensus where the dynamics are governed by the FJ model.

\textbf{Opinion manipulation in DW model.}
Pineda and Buend{\'{i}}a~\cite{Pineda2015} investigate the
effect of mass media in both DW and HK models. They consider
heterogeneous (in confidence radius) cases for both DW and HK
and study conditions under which the effect of mass media is maximized.
Another example of affecting
the network's opinion is presented in~\cite{Carletti2006}.

\textbf{Opinion manipulation in HK model.}
Standard tools in linear algebra enable one to understand
the dynamics of the DeGroot model. Such results help
to manipulate the opinion of the agents by modifying the topology
of the network~\cite{Dong2017}. However, this is not the only
way to manipulate the network's final state. 
Brooks and Porter~\cite{brooks2019model} use media 
to manipulate the outcome of the
discussion in a network; ``We maximize media impact in a 
social network by tuning the number of media accounts 
that promote the content and the number of followers of the accounts.''
We mentioned Hegselmann et al.~\cite{hegselmann2014optimal} 
extended the DeGroot model to include one strategic agent
to manipulate other agents. This paper also applies the same
idea to the HK model as well. Another work~\cite{hegselmann2015opinion} investigates the role
of a stubborn agent in an extension of the HK model. 
We include it in this section since it investigates 
the space of parameters to study conditions under which 
the stubborn agent (external signal) can maximize its 
influence in attracting other agents. 
They define \emph{intensity} of the signal as the 
number of times the signal is sent at time $t$. Interestingly, 
they discover higher intensities may have less effect in attracting
other agents. This result is similar to that of~\cite{Pineda2015}.
The intensity in case of~\cite{Pineda2015} is defined as 
the probability of interaction between a stubborn agent and another
agent in the DW model.

\textbf{Opinion manipulation in voter model.}
Gupta et al.~\cite{gupta2020} propose strategies
for manipulating the agents' opinion in a voter model.
The influence maximization on a complex network is given in~\cite{Moreno},
and influence maximization by considering the
agents' power of influence, the \emph{Influence Power-based Opinion Framework},
is proposed in~\cite{8750999}.

\textbf{More on opinion manipulation.}
We can mention Refs.~\cite{Hegselmann2015,rescu,Dietrich2018,9027368}
as other examples of opinion manipulation.
Goyal and Manjunath~\cite{9027368} build on~\cite{Aditya} and investigate 
a scenario in which two competing forces try to gain control
of the network and maximize the number of their followers. 
Each ``\emph{controller}'' has a budget constraint and Nash
control strategies are determined for each controller. Brede~\cite{Brede2019} 
investigates a \emph{rewiring} model in which \emph{influencers} 
try to maximize their impact.
\subsection{Power evolution}
We mentioned two properties of the final state in the DeGroot model.
If the network of agents is connected, i.e., the network does not consist of disjoint subgraphs,
then the final state is consensus and the consensus value is a weighted
average of the initial opinions. These two properties were the motivation for the introduction of
 power evolution in the DeGroot model.

\textbf{Power evolution in DeGroot model.}
Jia et al.~\cite{MirTabatabaei} study the evolution of power in a network. In the
 work of Jia et al.~\cite{MirTabatabaei}, the power of influence of agents
 evolves over a sequence of topics where the dynamic of each
 topic discussion is governed by the DeGroot model. 
 Suppose the network discusses
 a sequence of topics $s = 0, 1, 2, \cdots$ one after another where
the dynamic of each discussion is governed by the DeGroot model.
The weight matrix for each topic depends on the outcome of the previous topic:
\begin{equation}\label{eq:DeGrootFriedkin}
\mathbf{o}^{(t+1)}(s) = \mathbf{W}(s) \mathbf{o}^{(t)}(s)
\end{equation}
In Eq.~\ref{eq:DeGrootFriedkin} the weight matrix $\mathbf{W}(s)$
depends on the outcome of the topic $s-1$. 
This model is known as the DeGroot–Friedkin model.
The relation between social power and centrality ranking is established in~\cite{MirTabatabaei}.
Moreover, the conditions under which a democratic or autocratic structure is
formed are discussed. 

Kang et al.~\cite{KangDF} use a two-layer network to explore the evolution of social power
in the DeGroot-Friedkin model. Convergence properties of such a model are provided.
The weight matrix  $\mathbf{W}(s)$ is decomposed into a sum of two matrices
 $\mathbf{W}(s) = \mathbf{D(s)} + \mathbf{(I_n - D(s)) C}$ where $\mathbf{C}$ is called
 a \emph{relative influence matrix}. $\mathbf{C}$ is row-stochastic, irreducible with
 a zero diagonal. In the case of~\cite{KangDF}, there are two relative influence matrices,
 one for each layer. It is shown in~\cite{MirTabatabaei} that for the DeGroot-Friedkin model,
 a democratic configuration will be reached if and only if the relative matrix $\mathbf{C}$
 is doubly-stochastic. In the case of a two-layer network of~\cite{KangDF},
 the democratic configuration will be reached if both of the influence matrices are
 doubly-stochastic. They both have similar results for emerging autocratic configuration
 under star topology.

The evolution of individuals' power is 
further studied in~\cite{Friedkin2016socialPower,Jia2017,YeLatest,Ye2018,askarzadeh2,Askarzadeh2019,Tian2019}.
Ye and Anderson~\cite{YeLatest} extended the
DeGroot–Friedkin model by adding new characteristics to agents; 
``humbleness'' and ``unreactiveness.'' In this extension, power evolution of
agents is ``distorted'' by these new characteristics. These researchers study 
the existence and uniqueness of equilibria, 
 and convergence if it exists. 
Askarzadeh et al.~\cite{Askarzadeh2019} study the
power evolution of the DeGroot–Friedkin model using probability and Markov chain theory. 

\textbf{Power evolution in FJ model.}
Tian et al.~\cite{Tian2019} study power evolution in the FJ model, including the
properties of equilibria and the conditions under which democracy is achieved.
Furthermore, it is shown that  autocracy cannot be achieved in 
the presence of stubborn agents.

We mentioned earlier that stubborn agents are also called
media or leaders. The reason is that they can influence
other agents and have a major impact on the final
state of the system. Equivalently, stubbornness translates into
social power. This fact is not only observed in the FJ model, but also
in the DeGroot model (e.g.~\cite{Galam2016win}).
\subsection{Repulsive behavior}
The models we observed so far only support 
two types of behavior--attraction or indifference.
Humans are more complicated. If the topic is sensitive then
repulsive behavior emerges and causes polarization or fragmentation.
Let us look at models that support such behavior.

\textbf{Repulsion in the DeGroot model.}
Chen~\cite{backfire2019} adds a repulsive behavior to the 
model of Dandekar et al.~\cite{Dandekar5791}. The model
proposed in~\cite{backfire2019} uses a single parameter-- \emph{entrenchment parmeter}--
to capture both bias and \emph{backfire} effect. Their model also
supports polarization that previously did not exist in the original DeGroot model.

\begin{figure*}
    \centering
    \begin{subfigure}[b]{0.3\textwidth}
        \includegraphics[width=1\textwidth]{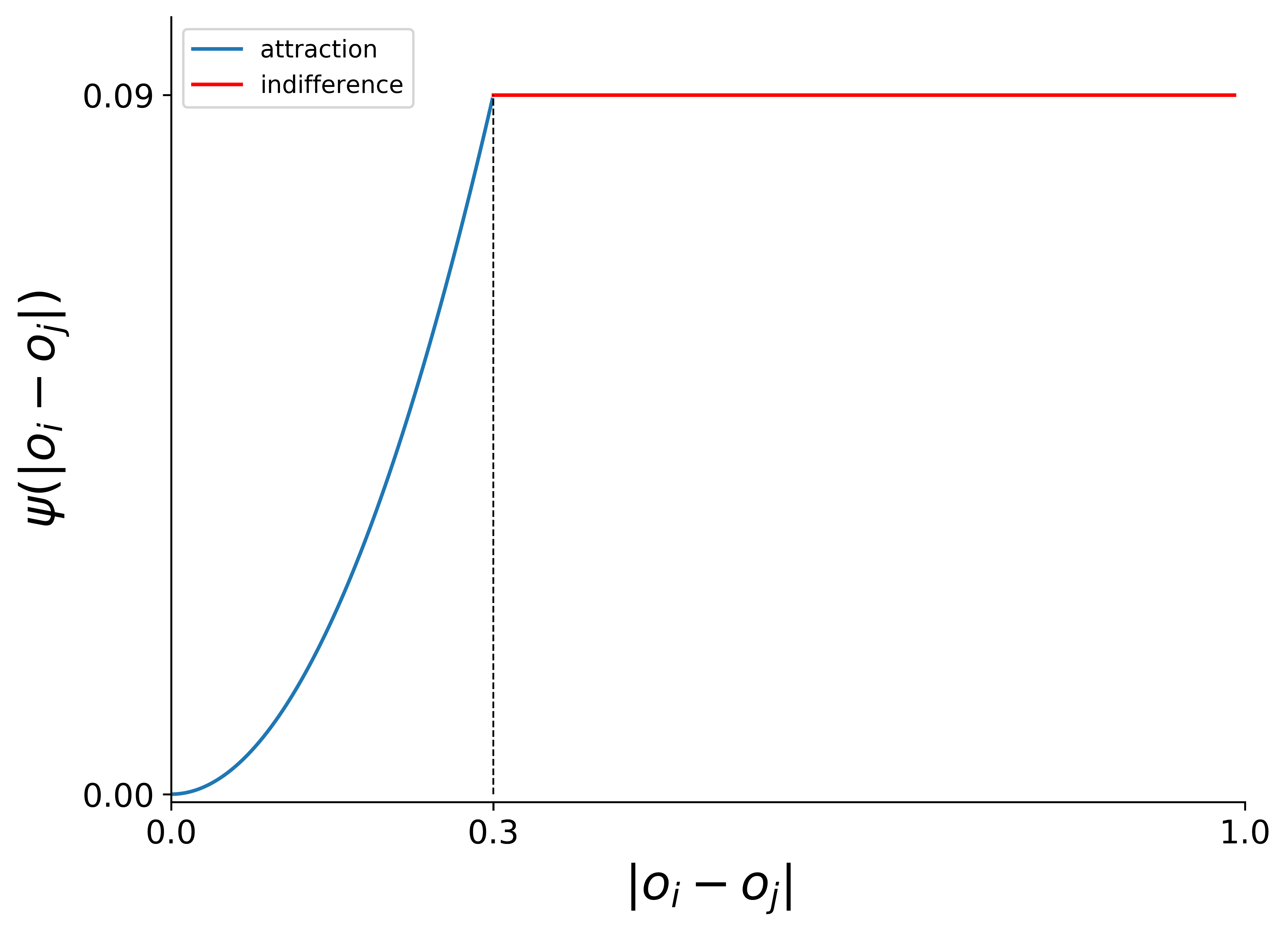}
        \caption{DW potential}
        \label{fig:DWPoten}
    \end{subfigure}
    ~
    \begin{subfigure}[b]{0.3\textwidth}
        \includegraphics[width=1\textwidth]{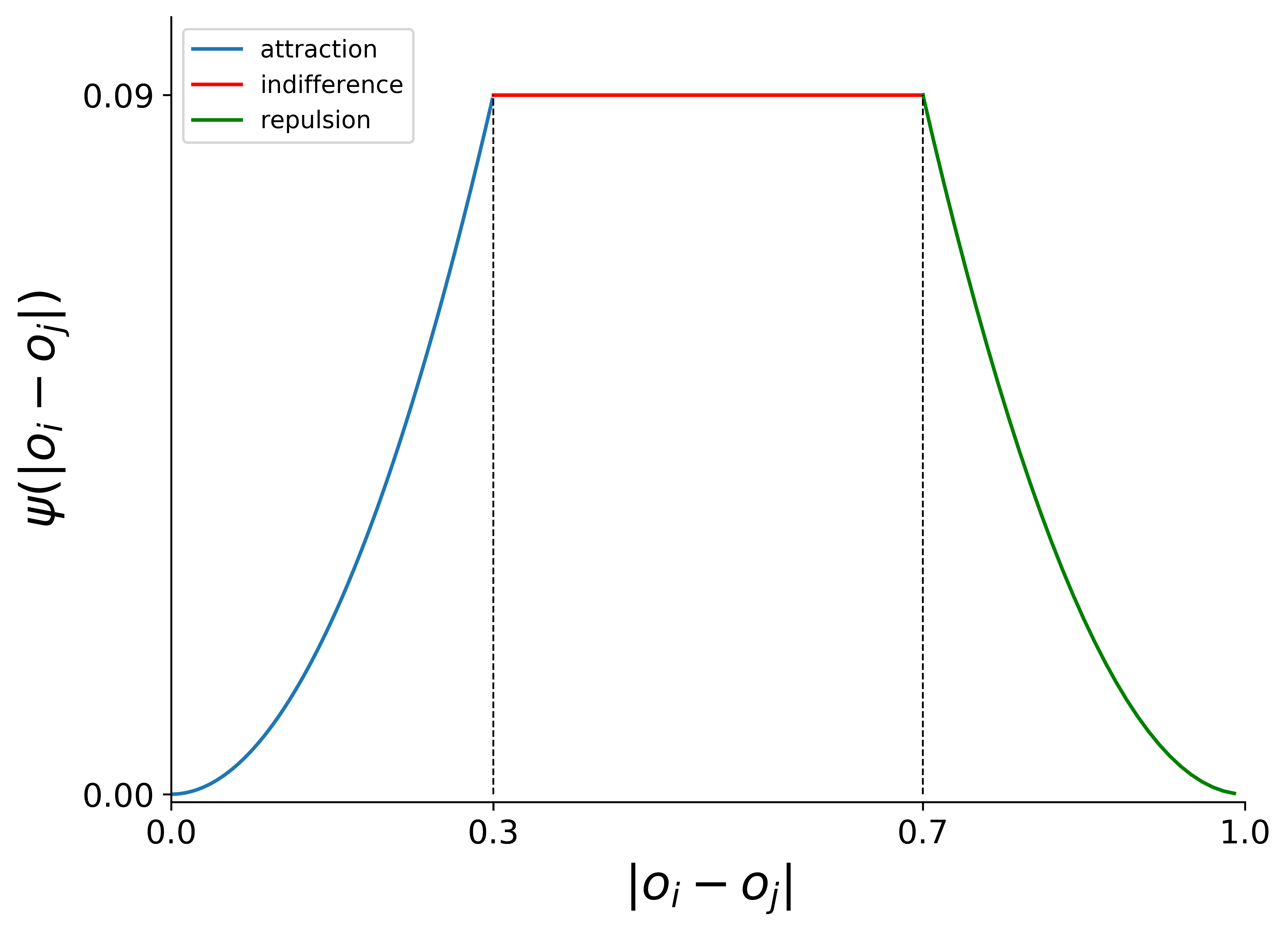}
        \caption{Jager-Amblard potential}
        \label{fig:JAPoten}
    \end{subfigure}
    ~ 
    \begin{subfigure}[b]{0.3\textwidth}
        \includegraphics[width=1\textwidth]{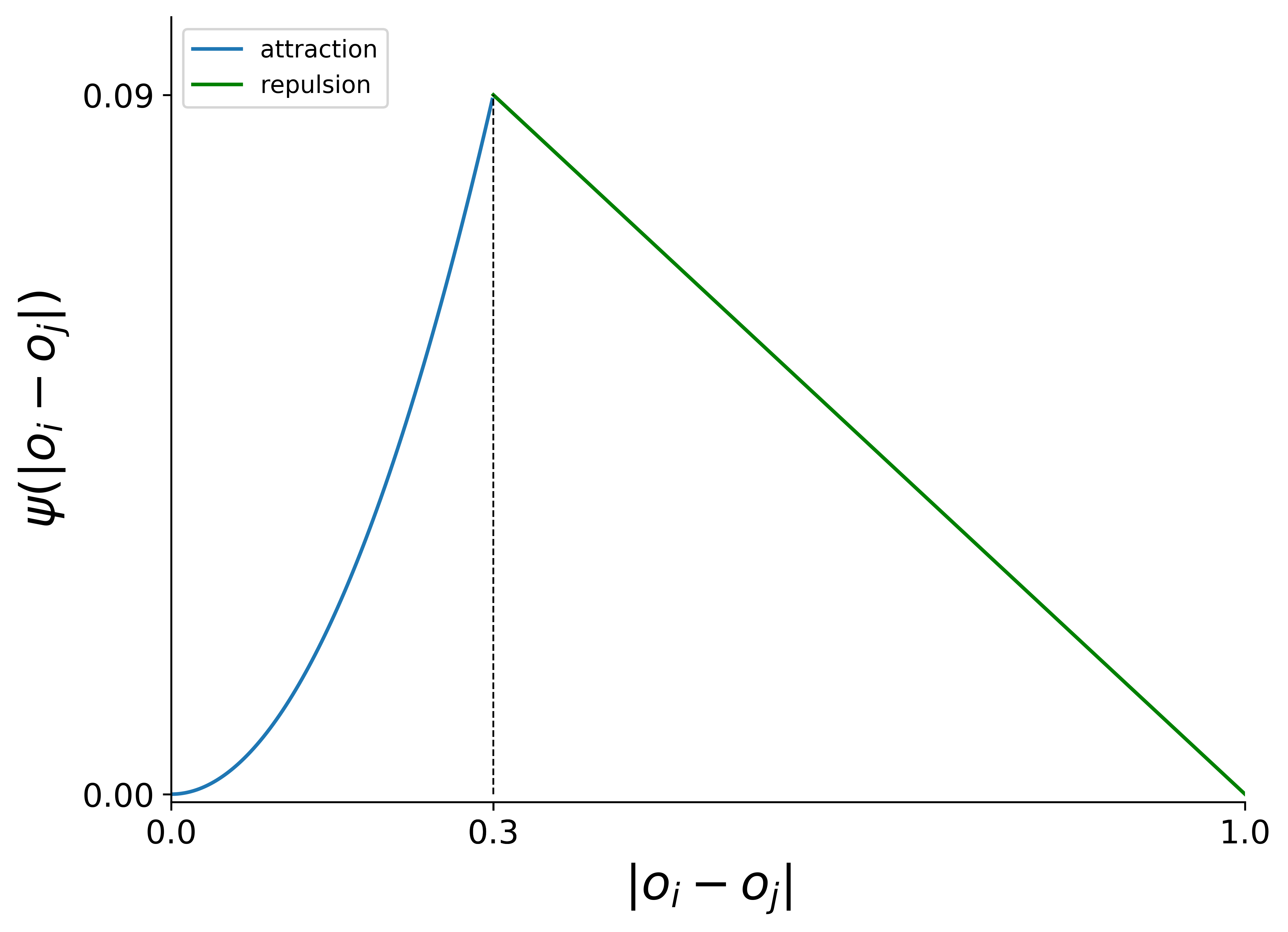}
        \caption{A repulsive DW potential}
        \label{fig:sfig2}
    \end{subfigure}
    \caption{Potential function examples. By choosing the potential function in (a) the model
    of~\cite{Noorazar2016} produces the DW model, and the potential function given in (b) results in Jager-Amblard model.  
The potential function in (c) generates another simple modified DW model with repulsive behavior.}
    \label{fig:example-potential}
\end{figure*}

\textbf{Repulsion in DW model.}
Repulsive behavior in a modified DW model is 
discussed in~\cite{kurmyshev2011,huet2008rejection,jager2005uniformity}. 
The model proposed in~\cite{Noorazar2016} is based 
on minimizing interaction energy between agents. 
The interaction energy is defined via \emph{potential functions}. 
The update rule in~\cite{Noorazar2016} is given by:
\begin{equation} \label{eq:updateNoWeights}
\left\{
  \begin{array}{lr}
    o_i^{(t+1)} &= o_i^{(t)}  -\frac{\mu}{2} \: \psi'(|o_i^{(t)} - o_j^{(t)}|) \: (o_i^{(t)} - o_j^{(t)})	\vspace{.1in}\\
    o_j^{(t+1)} &= o_j^{(t)}  +\frac{\mu}{2}  \psi'(|o_i^{(t)} - o_j^{(t)}|) \: (o_i^{(t)} - o_j^{(t)})
  \end{array}
\right.
\end{equation}
With the proper choice of a potential function, this model 
collapses to the DW model (see Fig.~\ref{fig:DWPoten}), or the model of 
Jager and Amblard~\cite{jager2005uniformity} (see Fig.~\ref{fig:JAPoten}). 
Using the potential function in Fig.~\ref{fig:JAPoten}, the model will support
three types of behavior: attraction, indifference, and repulsion.
The potential function given in Fig.~\ref{fig:sfig2} produces a modified DW model
with repulsive behavior. In  these three potential functions the confidence radius is
$r = 0.3$.

\textbf{More on repulsive behavior.}
 References~\cite{Altafini2012,Altafini2013,Altafini2018,Schweighofer,Proskurnikov2016,bhat2019opinion,HE2019,Zhang2014Control,Meng2019,Dehghani2020}
consider signed graphs and the concept of balance theory~\cite{Cartwright1956} in their work
of modeling antagonistic or repulsive behavior. In a signed graph each edge
is labeled with a positive or negative sign, defining 
friendship or antagonistic relationships.
Aghbolagh et al.~\cite{Dehghani2020} has implemented
three types of behavior--attraction, indifference, and repulsion--in
a modified HK model. They show
their new model can lead to consensus, 
bipartite consensus, and clustering of opinions.
\subsection{Noisy models}
Noise is injected into models
for different purposes. For instance, M{\"{a}}s~\cite{Mas2010} uses
noise to implement the idea of the \emph{tendency for uniqueness} in 
the model of Durkheim~\cite{Durkheim}. The strength of 
noise in this model increases as the clusters grow in size.
Other forms of noise are implemented to model 
different traits of humans' behavior. Noise can be used to model the
death or birth of an agent, and to mimic internal thoughts
or interactions with external sources such as media or
 books. 
Below we review some of the noisy models.

\textbf{Uncertainty in DeGroot model.}
A modified DeGroot model, taking into account uncertainty of 
agents encoded as intervals is given in~\cite{kurmyshev2011}.

\textbf{Noise in DW model.}
One can argue humans do not have a 
sharp threshold like a confidence
radius for accepting or rejecting other ideas. Grauwin and Jensen~\cite{Grauwin2012} 
use random noise in the DW model to kill the aforementioned sharp threshold.
in~\cite{Grauwin2012}, two agents interact with a probability that depends
on the difference of opinion of the two agents--an \emph{interaction noise}. Another type
of noise introduced in~\cite{Grauwin2012} is reminiscent of the death of a person and birth of
another; an agent changes its opinion at time $t$ to a random opinion with some probability.
Pineda et al.~\cite{Pineda2011} investigate another type of noise in the DW model:
``Individuals are given the opportunity to change their opinion, with a given probability, to a
randomly selected opinion inside an interval centered around the present opinion.''
References~\cite{Carro2013, Quattrociocchi,Baccelli2017,zhang2018robust,pineda2009noisy} 
also investigate the noise effect in models inspired by the DW model.

\textbf{Noise in HK model.}
Su et al.~\cite{Su2017} introduce noise
to the homogeneous HK model and show how it can help the formation of consensus.
A recent study investigates the role of \emph{environment and communication noise}
in the heterogeneous HK model~\cite{8918332}. 
Phase transition and convergence time are studied
in the presence of environmental noise in~\cite{8918332}.
Another example of noise in the HK model is discussed in~\cite{Pineda2013}.
Nonlinear stability for the HK model in the 
presence of noise is addressed in~\cite{Chazelle2017365}.
A modification of the HK~\cite{Liang2016} models uncertainty of agents.
In this model, some agents may have 
an opinion that is actually an interval, not a 
single number. 

\textbf{Noise in Galam model.}
Hamann~\cite{HamannNoise} explores noise in a modified Galam model
in a group of mobile agents with the presence of contrarians.

\textbf{Noise in Sznajd model.} 
Sabatelli and Richmond~\cite{Sabatelli} add noise to a modified Sznajd model
where the updates are done in a synchronous fashion. 
One of their results in that they `` predict that consensus can be increased by the 
addition of an appropriate amount of random noise.''

\textbf{Noise in voter model.} 
One of the earliest noisy voter models is~\cite{GRANOVSKY199523},
which employs standard statistical physics techniques and
examines the critical behavior of the system and its phase transition.
References~\cite{carro2016noisy,Peralta2018} examine
the role of noise in the voter model on complex networks. 
The role of ``zealots'', i.e., fully-stubborn agents, in a noisy voter model
is inspected in~\cite{PhysRevEKhalil} where agents form a fully connected graph.
\subsection{Interrelated topics}
It is rarely the case that a given issue exists in 
an isolated environment. The change of opinion about one
topic could cause a change of opinion about another topic.
For instance, a change of opinion about health
can lead to a change of opinion about exercise and diet. 
This area has not yet seen much exploration. Below,
we present the limited models of this nature.

\textbf{Interrelated topics in FJ model.}
In 2016, two independent works~\cite{Noorazar2016,Friedkin321} 
proposed novel ideas for the dynamics of interrelated 
(coupled or interdependent) topics. The model
in~\cite{Noorazar2016} is novel and does not
fall under the umbrella of classical models; however, the Ref.~\cite{Friedkin321} 
is a generalization of the FJ model.
Friedkin et al.~\cite{Friedkin321} revisit the idea of interdependent topics
in the multidimensional FJ model in~\cite{Parsegov7577815}.
Tian and Wang~\cite{TIAN2018213} 
introduce the idea of sequentially dependent topics
in which each topic is discussed in a sequence and the
outcome of topic $s$ affects the dynamics of the discussion of
topic $s+1$ where each topic's dynamic is governed by
the FJ model.

\textbf{Interrelated topics in DW model.}
Fei et al.~\cite{Fei2017} propose 
a model for interdependent topics where interactions are 
pairwise and follow the bounded confidence concept.

\textbf{More on interrelated topics.}
Ahn et al.~\cite{9018151} propose a novel opinion model with interrelated
topics. Let $\ell$ and $s$ to be two given coupled topics. 
in~\cite{Noorazar2016}, for example, the interaction between agents is based on a given topic. 
Suppose agents $i$ and $j$ interact about topic $\ell$ and update
their opinion. Consequently, by internal thoughts due to the coupling of 
$\ell$ and $s$, their opinion about $s$ will be updated as well, despite
the fact that they did not discuss  topic $s$. However,
in~\cite{9018151} it is possible that topic $\ell$ of agent $i$ is coupled
with topic $s$ of agent $j$. Other novel models have recently been proposed, 
such as~\cite{9003276}, \emph{Bayesian learning} model~\cite{GengBayesian}, 
a model based on \emph{Achlioptas Process}
\cite{wang2019opinion} and a model based 
on \emph{Latan\'{e}'s social impact theory}~\cite{kowalskastycze},
to name a few.
\subsection{Expressed vs. private opinions}
The expressed opinion of
agents is not always identical to their true internal belief. Social pressure
can cause people to express an opinion that
 aligns with that of others while contradicting their internally held belief. 
This concept was proposed in 1990 by Nowak and Latan{\'e}~\cite{nowak1990private}
which is based on earlier work of Latan{\'e}~\cite{latane1981psychology}.
In this section, we present more recent models that consider 
the co-evolution of expressed and personal opinions.

\textbf{Expressed vs. private opinions in FJ.}
The dynamics of the co-evolution of expressed and 
private opinions (EPO) in the FJ model along with its convergence properties are 
presented in~\cite{ye2019_EPO}. 

\textbf{Expressed vs. private opinions in voter model.}
References~\cite{Gastner2018,10poneEP,HeterogeneousGibert} explore different
ideas about expressed and private opinions in the voter model.

\textbf{More on EPO.}
Some novel models explore the dynamics
of private and expressed opinions~\cite{ShengWen,huang2014,Medina2019, 8931758};
while they do not fall under the umbrella of well-known models, they are worth mentioning.

We mentioned agents may reveal an opinion that is different from
their internal true belief. in~\cite{ye2019_EPO} 
the co-evolution of the two (internal and revealed) opinions are studied.
We have also talked about manipulating the agents to influence them toward a predetermined
target. The model presented by Afshar and Asadpour~\cite{Afshar2010} is somewhere in between.
Their model is inspired by the DW model and includes some \emph{informed} agents who
pretend their opinion is close to that of other agents. These informed agents influence other agents toward
a predetermined target opinion.

Table.~\ref{tab:overviewTbl} provides an overview of the material presented in this paper.
\section{Last words}\label{LastWords}
Before closing the discussion, we would like
to cover other interesting models that have not yet been studied extensively 
 and acknowledge
the great efforts of other researchers.\\

Li et al.~\cite{Li2017} propose an interesting model. 
Unlike BCMs, agents in their
model of~\cite{Li2017} interact if the difference of opinion is larger 
than a threshold due to social pressure. In the model given 
in~\cite{LiuQipeng} there is a potential to interact with 
agents whose opinions outside of the confidence interval.

Zhang and Hong~\cite{Zhang2013} propose two synchronous 
versions of BCM in which not all neighbors
of agent $i$ participate in the update of the opinion of agent $i$.
Instead, several neighbors are selected randomly. 
These researchers are interested in the convergence
properties of this model, which sits between the pairwise interaction
in the DW model and the synchronous HK model. 
In this model, there is a potential to interact
with agents beyond one's confidence radius. More details about the model are given in~\cite{Zhang2019Plos}. 
Another interesting work~\cite{Chazelle2017}
studies convergence of a modified HK model in which 
agents have \emph{inertia}. References~\cite{Fu2015,schawe2020open,Choi2019} 
contain other examples of the study of the convergence 
properties of the HK model and its variations. Gang et al.~\cite{Kou2012} 
investigate the final state of the heterogeneous (in confidence radius) HK model.

Rubio et al.~\cite{9016083} have recently proposed a model for anomaly 
detection in the Industrial Internet of Things architectures.
Other examples of applications of opinion dynamics in engineering are given in~\cite{kuhn2018population}
and~\cite{Pilyugin}, where the later reference studies voting processes 
inspired by BCM.\\

Physics has inspired different models of opinion dynamics, of course.
There are several works based on the kinetic theory of 
gases~\cite{Helbing1993,toscani2006kinetic,Boudin2009,Biswas2012,Pareschi2017,alexanian2018anti,oestereich2020hysteresis,Lachowicz,fraia2020boltzmann,lachowicz2019diffusive,lima2019kinetic}
where interactions are defined by Boltzmann type equations. 
Applications of such models in other fields, such as economics, are 
found in the book by Pareschi and Toscani~\cite{pareschi2013interacting}.
D{\"u}ring et al.~\cite{During2009} investigate the presence of 
leaders and Wang et al.~\cite{Wang2017} took the
effect of noise into account in these dynamics. Furthermore, the mean-field theory has been 
 employed to explore the 
landscape of dynamics~\cite{Biswas2011,Chowdhury2011,pareschi2019mean,wang2020robust}. 
The Ising model is another tool that can be used when the 
opinion space is binary~\cite{chmiel2018q,bottcher2017critical,Biswas2017,biswas2009new},
with applications in areas such as group decision making~\cite{galam1997rational,Sznajd2000}.\\

Lastly, it is worthwhile mentioning 
the evolution of agents' \emph{susceptibility to persuasion}, which is examined in~\cite{Abebe2018,Chan2019}. 
Edge weights are used to implement the frequency of interaction between agents  in~\cite{Patterson,Noorazar2017}. For  more 
details about continuous-opinion-space models, we refer
the reader to the tutorials in Refs.~\cite{noorazar2019classical,Proskurnikov2018}.
\section{New questions}
While a great deal of progress has been made in the field, there is still
great potential for improvement. Humans do not interact with
all their neighbors simultaneously, unlike the DeGroot model. Even for
a network of computers that can interact quickly and can follow a clear
set of rules, there are physical limitations.
Balanced graphs are used to model repulsive behavior based
on principles such as ``friend of my friend, is my friend'' or 
``enemy of my enemy is my friend,'' which are not always true.

While some of the opinion dynamics models are designed to 
model a certain trait (e.g., homophily)
or are tailored to create interesting dynamics (e.g., preventing consensus by introduction of noise), 
these models are not universal. Hence, it would be interesting to shrink 
the gap between simplicity of theoretical models and complexity of
humans' behavior. 

Opinion dynamics could be used to detect
fake-news resources on social media. Detecting susceptible
individuals who might be attracted to terrorist groups via the Internet
is another potential domain of work.

It would be helpful to see more applications of opinion
dynamics in real world problems. More specifically, it would 
be fascinating to take advantage of opinion dynamics models to detect and flag computers
or processors sending erroneous or corrupted messages in computer buses.
\begin{center}
    \addtolength{\tabcolsep}{-0.3pt}
    \sisetup{group-digits=integer}
    \begin{tabular}[]
    {
            @{}
            l
            l
            l
            @{}
        }
        \hline
        Model   &   {Objection} &   {References} \\
        \hline
        DeGroot   &   {Convergence}                 &   \cite{Berger1980}   \\
                         &  {Stubbornness}                 & \cite{FJmodel1990,Wai2016,abrahamsson2019,ZHOU2020363} \\
                         &  {Bias}                                & \cite{Dandekar5791,xia2019analysis}  \\
                         &  {Opinion manipulation}     & \cite{ZHOU2020363,Dong2017,Pineda2015,Bauso2018}  \\
                         &  {Repulsion}                       & \cite{backfire2019}  \\
                         & {Power evolution}              & \cite{MirTabatabaei,KangDF,Friedkin2016socialPower,Jia2017,YeLatest,Ye2018,askarzadeh2,Askarzadeh2019,Tian2019} \\
\hline
FJ               & {Convergence}                     & \cite{Parsegov7577815} \\
                   & {Opinion manipulation}         & \cite{gaitonde2020} \\
                   &{Power evolution}                 & \cite{Tian2019} \\
                   & {Interrelated topics}              & \cite{Friedkin321,Parsegov7577815,TIAN2018213} \\
                   & {EPO}                                   & \cite{ye2019_EPO} \\
\hline  
DW             & {Convergence}                     & \cite{Fortunato2004,lorenz2010heterogeneous,Chen2019,zhang2015conv,Shang2014,Huang2018} \\
                   & {Stubbornness}                    & \cite{huang2016modeling} \\
                   & {Bias}                                   & \cite{Sirbu2019} \\
                   & {Opinion manipulation}        & \cite{Pineda2015} \\
                   & {Repulsion}                          & \cite{kurmyshev2011,huet2008rejection} \\
                   & {Interrelated topics}             & \cite{Fei2017} \\
                   & {Noise}                                 & \cite{Grauwin2012,Pineda2011,Carro2013, Quattrociocchi,Baccelli2017,zhang2018robust,pineda2009noisy} \\
\hline
HK             & {Convergence}                     & \cite{Hegselmann2002,Bhattacharyya}  \\
                   & {Stubbornness}                   & \cite{brooks2019model,hegselmann2015opinion} \\
                   & {Bias}                                  & \cite{Chen2017} \\
                   & {Opinion manipulation}       & \cite{brooks2019model,hegselmann2014optimal,hegselmann2015opinion} \\%
                   & {Noise}                                & \cite{Su2017,Pineda2013,Chazelle2017365,8918332} \\
\hline
Galam         & {Convergence}                  &  \cite{GortnerBernd} \\
                   & {Stubbornness}                  & \cite{GalamInflex,CHEON20181509,galam2019tipping,QIAN2015187,CHEON2016429}  \\
                   & {Contrary}                          & \cite{GalamAsymmetric,GalamInflex} \\
                   & {Noise}                               & \cite{HamannNoise} \\
\hline
Voter         & {Stubbornness}                  & \cite{mukhopadhyay2020,PhysRevEKhalil,Mobilia2007,yildiz2013binary} \\
                 & {Opinion manipulation}       & \cite{gupta2020,Moreno,8750999} \\
                 & {EPO}                                 & \cite{Gastner2018,10poneEP,HeterogeneousGibert} \\
                 & {Noise}                                & \cite{GRANOVSKY199523,carro2016noisy,PhysRevEKhalil,Peralta2018} \\
        \hline
    \end{tabular}
    \captionof{table}{Overview of materials presented in the paper.}
    \label{tab:overviewTbl}
\end{center}

\section{Conclusions}\label{SecConcl}
In this paper, we reviewed well-known models
of opinion dynamics for both
continuous and discrete opinion spaces. In the continuous-opinion-space case, 
we reviewed the DeGroot model, and one of its major extensions, namely the FJ model.
Afterwards, we presented the two major bounded confidence models,
namely the DW model and the HK model. In the discrete-opinion-space 
case, we reviewed the Galam model, the
Sznajd model and the voter model. 
Subsequently, for the selected additional models, reviewed some extensions
that added extra ingredients(s) to the original model--
stubbornness, bias, repulsive behavior, power evolution, 
interrelated topics, noise, and expressed and private opinions. Finally, 
we posed new questions for future explorations.
\section{Acknowledgments}
We would like to acknowledge the insightful inputs of Rainer Hegselmann and Mohammad Hossein Namaki
that immeasurably helped in the development of this manuscript.

This is a pre-print of an article published in The European Physical Journal Plus. The final authenticated version is available online at: \href{https://doi.org/10.1140/epjp/s13360-020-00541-2}{https://doi.org/10.1140/epjp/s13360-020-00541-2}.

\bibliographystyle{unsrt}
\bibliography{2020_survey_refs}
\end{multicols}
\end{document}